\documentclass[twocolumn,pra]{revtex4}
\usepackage[mathscr]{eucal}
\usepackage{amsmath}
\usepackage{amssymb}
\usepackage{amsfonts}
\usepackage{ae}
\usepackage{epsfig}

\def\dd{\mathrm{d}}
\def\r){\right)}
\def\l({\left(}

\begin{document}

\title{Image transmission through a stable paraxial cavity}

\author{Sylvain Gigan, Laurent Lopez, Nicolas Treps, Agnès Maître, Claude Fabre}

\affiliation{Laboratoire Kastler-Brossel, Université Pierre et Marie Curie, Case 74, 75252 PARIS cedex 05.} 

\email{gigan@spectro.jussieu.fr}

\begin{abstract}
We study the transmission of a monochromatic "image" through a
paraxial cavity. Using the formalism of self-transform functions,
we show that a transverse degenerate cavity transmits the
self-transform part of the image, with respect to the field
transformation over one round-trip of the cavity. This formalism
gives a new insight on the understanding of the behavior of a
transverse degenerate cavity, complementary to the transverse mode
picture. An experiment of image transmission through a
hemiconfocal cavity show the interest of this approach.
\end{abstract}

\maketitle 

\section{Introduction}

Image transmission and propagation in a paraxial system, using
optical devices such as lenses and mirrors is a well-known and
extensively studied problem \cite{BornWolf}. The free-propagation
of a field changes its transverse distribution, but in some
planes, such as conjugate planes, or Fourier plane, one get simple
transformations of the image. On the other hand, transmission
through cavity has a drastic effect on the transverse distribution
of the field, as one must take into account the transverse
characteristics and the resonances of the cavities. Optical
cavities have also been studied extensively for a long time,
starting from the Fabry-Perot resonator, then to the
laser\cite{Siegman}, and their are commonly used as temporal
frequency filters. Less known are their spatial frequency filter
properties. An optical cavity is associated to an eigenmode basis,
i.e. a family of modes (like $TEM_{pq}$ hermite gaussian modes)
which superimpose onto themselves after a round-trip inside the
cavity. This basis depends on the geometrical characteristics of
the cavity (length, curvature of the mirrors, ...). Only
eigenmodes can be transmitted through the cavity at resonance and
the cavity acts both as a spatial filter and frequency filter.
This mode selection property of cavities, that does not exist in
free propagation, is well-known in the longitudinal domain for
frequency filtering. However, the general transverse effect of a
cavity on an image has, to the authors' knowledge, never been
carefully investigated. Whereas the transmission of an image
through a cavity which is only resonant for one transverse mode is
well-known to be completely destructive for this image, some
particular cavities called transverse degenerate cavities can
partially transmit an image, in a way that we will precise in the
present paper.

This work is part of a more general study on quantum effects in
optical images \cite{Kolobov:95,Kolobov:99} and more precisely on
noiseless parametric image amplification \cite{Kumar99, Lantz00},
performed in the continuous-wave regime. In order to have a
significant parametric gain with low-power laser, we need resonant
optical cavities, operating in the regenerative amplifier regime,
below, but close to, the oscillation threshold \cite{Kimble}. As a
first step, we therefore need to precisely assess the imaging
properties of an empty optical cavity. This study turns out to be
interesting in itself, and might also be useful for other
experiments.

We begin this paper by reminding in section II some useful
features of paraxial propagation of an image and of degenerate
cavities. In section III, we develop a new formalism to understand
the transmission of an image through a paraxial cavity, and link
it to the formalism of cyclic transforms. In section IV, we show
simulations and experimental results of image transmission through
a simple degenerate cavity : the hemi-confocal cavity.

\section{"ABCD" cavity round-trip matrix transforms}

All the theory developed in this paper will be performed within
the paraxial approximation. We consider a monochromatic
electromagnetic field E($\vec{r}$,t) at frequency $\omega$,
linearly polarized along a vector $\vec{u}$ and propagating along
a direction $z$ of space. The position in the transverse plane
will be represented by the vector $\vec{r}= x \vec{i}+y\vec{j}$.
The electric field is supposed stationary and can be written in a
given transverse plane as :
\begin{equation}
 \vec{E}\left(\vec{r},t\right)= Re [E \left(\vec{r}\right)  e^{-i \omega
 t}\vec{u}]
 \end{equation}
where $\vec{u}$ is the polarization unit vector. The local
intensity in this plane is then :
\begin{equation}
I\left(\vec{r}\right) = 2 \epsilon_0 c E\left(\vec{r}\right)
E^*\left(\vec{r}\right) .
\end{equation}

The input image considered all along this paper is defined by a
given transverse repartition of the complex transverse field
$E_{in}(\vec{r})$ in an input plane $z_{in}$. We suppose that its
extension is finite around the optical axis, and that its
transverse variations are such that this image propagates within
the paraxial approximation. We will consider both intensity images
and "field" images, i.e. not only the transverse intensity
distribution of the field, but also the amplitude distribution
itself.

\subsection{Image propagation in a paraxial system}

The field $E(\vec{r})$ is propagating through an optical system
along the $z$ axis. An input-output relation of the form can be
written :
\begin{equation}
E_{out}\left(\vec{r}\right) = \mathcal{T}
[E_{in}\left(\vec{r}\right)]
\end{equation}
where  $E_{in}(\vec{r})$ and $E_{in}(\vec{r})$ are the fields just
before and just after the optical system and $\mathcal{T}$ is the
transformation of the field associated to the optical system. If
the system is only a linear paraxial system (made of lenses, or
curved mirrors, but without diaphragms), the propagation
properties of the system are described by its Gauss matrix T
(often called ABCD matrix) which writes :
\begin{equation}
T=\begin{pmatrix} A & B \\
C & D
\end{pmatrix}
\end{equation}

All the properties of the system can be inferred from the values
of the coefficients A, B, C and D, and of the total optical length
$L$ of the system (responsible for a phase factor which is not
included in the ABCD coefficients). We will assume that the index
of refraction is the same at the input and at the output of the
system. As a consequence, we have $\det(T)=AD-BC=1$. In
particular, the transformation $\mathcal{T}$ of the field can be
derived from the Huygens-Fresnel equation in free space in the
case $B\neq 0 $ \cite{Siegman}:
\begin{eqnarray}\label{HuygensParaxial}
\mathcal{T} : E(\vec{r_1}) & \rightarrow  & E(\vec{r}_2)= - e^{i k
L}\frac{i}{B \lambda}\iint \dd^2 \vec{r}_1 E(\vec{r}_1)
\\\nonumber && \exp\left[{ -i \frac{\pi}{B \lambda}\l( A
{\vec{r}_1}^2-2 \vec{r}_1 \vec{r}_2 + D {\vec{r}_2}^2 \r) }\right]
\end{eqnarray}

If $B=0$, the Gauss matrix can be written
 $T=\begin{pmatrix}
  M & 0 \\
  C & \frac{1}{M}
\end{pmatrix}$.
In this case the field in the output plane is given by:
\begin{eqnarray}\label{HuygensParaxial2}
\mathcal{T} : E(\vec{r}_1)\rightarrow E(\vec{r}_2) &=& -e^{i k L}
M u_1(M \vec{r}_1) e^{\frac{i k C M \vec{r}_2^2}{2}}
\end{eqnarray}

In terms of imaging, a conjugate plane corresponds to a
transformation for which one retrieves the input image within a
magnification factor M. From equations (\ref{HuygensParaxial}) and
(\ref{HuygensParaxial2}), it can be inferred that:
\begin{itemize}
\item if $B=0$ one retrieves the intensity image
 but not the amplitude (there is a phase curvature coming
from the term $e^{\frac{i k C M \vec{r}_2^2}{2}}$ of equation
(\ref{HuygensParaxial2})). We will call such a transform an
"Intensity-Conjugate Transform", or ICT. \item if $B=0$ and $C=0$
one retrieves the amplitude image (and the intensity image of
course). We will call such a transform an "Amplitude-Conjugate
Transform", or ACT. This transform is sometimes also called a
Near-Field (NF).
\end{itemize}

Another interesting transformation is the one for which one
obtains the spatial Fourier transform of the image. Still from
equations (\ref{HuygensParaxial}) and (\ref{HuygensParaxial2}),
one sees that:
\begin{itemize}
\item if $A=0$ one obtains the Fourier transform for the field,
within a curvature phase term corresponding to the factor
$e^{\frac{-i \pi D \vec{r}_2^2}{B \lambda}}$ of equation
(\ref{HuygensParaxial})). This factor does not affect the
intensity distribution. We will call this transformation a
"Intensity Fourier Transform", or IFT.

\item if $A=0$ and $D=0$ one obtains a perfect spatial Fourier
transform for the amplitude field. We will call this
transformation an "Amplitude Fourier Transform", or AFT. It is
sometimes called a far-field (FF).
\end{itemize}

It is straightforward to see that a 2f-2f system (a lens of focal
distance $f$ placed before and after a distance $2f$) performs an
ACT, and that a f-f system performs an AFT. Whereas AFT and ACT
can be simply and directly juxtaposed side-by side, this is not
the case for IFT and ICT transformations because of the phase
factors.

Let us remind a few obvious facts, which will be nonetheless
useful to understand the rest of the discussion. Two length scales
have to be considered for the optical system length L : The "rough
length", important to understand propagation (diffraction)
effects, and the "exact length", which must be known on the scale
of $\lambda$, necessary to determine the exact phase of the field.

\subsection{Transverse degeneracy of a resonator}

For simplicity purposes, all our discussion about cavities will be
restricted to the case of linear optical cavities with two
spherical mirrors. Its extension of the discussion to more complex
cases (ring cavity, cylindrical mirrors, etc...) is
straightforward. We also assume that the transverse extension of
the field is not limited by the size of the mirrors. In this
simple case the cavity is fully described by its round-trip Gauss
matrix $T_{cav}$, starting from a given reference plane.

We consider here only geometrically stable cavities ($|A+D|>2$).
In this case, the eigenmodes of the device are the Hermite-Gauss
modes (HG) adapted to the cavity, i.e having wavefront coinciding
with the mirror surfaces. The normalized transverse electric field
in the $TEM_{mn}$ mode basis is given by:
\begin{eqnarray}\label{Amn}
  A_{mn}(\vec{r},z) &=&  C_{mn} \frac{1}{w(z)} H_m\l(\frac{\sqrt{2}x}{w(z)}\r)
H_n\l(\frac{\sqrt{2}y}{w(z)}\r)\nonumber\\ && e^{ik\l(\frac{r^2}{2
q(z)}\r)} e^{-i(n+m+1)\arctan\l(\frac{z}{z_R}\r)} e^{i k z}
\end{eqnarray}

where:
\begin{eqnarray}
C_{mn} &=& \frac{1}{\sqrt{\pi 2^{m+n-1}m!n!}}\nonumber\\
z_R  &=& \frac{\pi w_0}{\lambda}\nonumber\\
q(z) &=& z - i z_R\\
w(z) &=& w_0 \sqrt{1+\l(\frac{z}{z_R}\r)^2}\nonumber\\
\Psi(z) &=& (n+m+1) \arctan\l(\frac{z}{z_R}\r)\nonumber
\end{eqnarray}

$w_0$ is the waist of the $TEM_{00}$ mode of the cavity taken in
its focal plane, of coordinate $z=0$, and $q$ is the complex
radius of curvature. It is important to note that $q$ is
independent from $m$ and $n$, and only depends on the position and
size of the waist. Finally, $\Psi(z)$, the Gouy phase-shift, will
play a major role in this discussion.

Let us note $z_1$ and $z_2$ the mirror positions and $L=z_2-z_1$
the total length of the cavity. The resonant cavity eigenmodes will be the
HG modes $A_{mn}$ having a total round-trip phase-shift is equal
to $2p'\pi$, with $p'$ integer. If the input field has a fixed
wavelength $\lambda$, this will occur only for a comb of cavity
length values $L_{mnp'}$ given by:

\begin{equation}\label{numnp}
L_{mnp'} =  \frac{\lambda}{2}\l(p'+(n+m+1)\frac{\alpha}{2 \pi}\r)
\end{equation}

where

\begin{equation}
\alpha=2
\left(\arctan\left(\frac{z_2}{z_R}\r)-\arctan\l(\frac{z_1}{z_R}\right)\right)
\end{equation}
is the Gouy phase shift accumulated by the $TEM_{00}$ mode along
one cavity round-trip. It is related to the cavity Gauss matrix
$T_{cav}$ eigenvalues $\mu_{12}$ by the relation:
\begin{equation}
\label{ABCDGouy}
\mu_{1,2} = e^{\pm i \alpha}
\end{equation}

This simple relation has been shown in \cite{dingjan}  for a linear
cavity with two curved mirrors. We give in the appendix a
demonstration of this result valid for any stable paraxial cavity.

A cavity is called "transverse degenerate" when for a given
frequency and cavity length, several transverse modes are
simultaneously resonant. From equation (\ref{numnp}), we can see
that:
\begin{itemize}
\item there is a natural degeneracy for HG modes giving the same
value to $s=m+n$, related to the cylindrical symmetry of the
device. We will not consider this natural degeneracy any longer,
and call s the transverse mode order and p the longitudinal mode
order;
 \item the cavity is transverse degenerate when $\alpha$ is a rational
fraction of $2\pi$. Let us write $\alpha/2\pi$ as an irreducible
fraction:
\begin{equation}\label{KN}
  \alpha   = 2 \pi \frac{K}{N} [2\pi]\mbox{.}
\end{equation}
with $K$, $N$ integers and $0 <\frac{K}{N}< 1$. $K/N$ is called
the \textit{degeneracy order of the cavity}\cite{dingjan}.
\end{itemize}

As the degeneracy order is the remainder part of the total Gouy
phase-shift over one turn of the cavity, we conclude that there
exists an infinite number of cavity configurations with the same
degeneracy order. Furthermore, the rational fraction ensemble
being dense in $\mathbb{R}$, transverse degenerate cavities is a
dense ensemble among all the possible cavities.

\begin{figure}[htbp]
\begin{center}
\epsfig{file=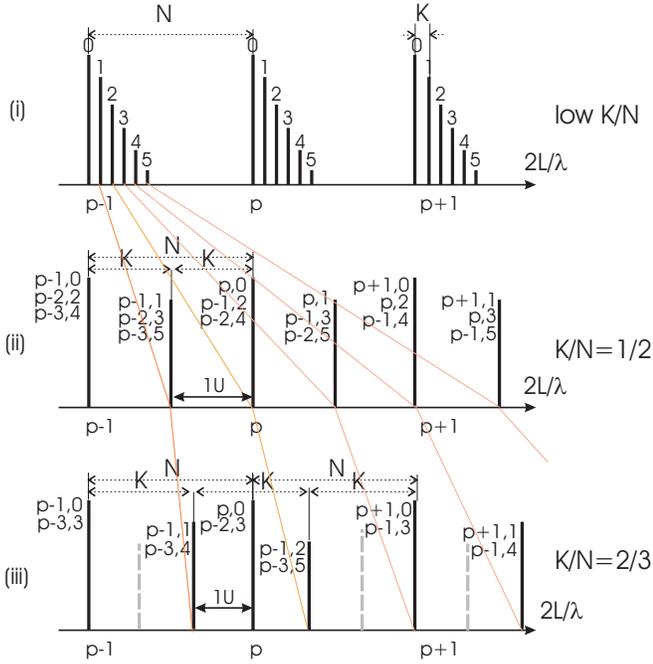,width=1\linewidth} \caption{partial
transverse and longitudinal comb in 3 configurations. (i) low  K/N
cavity (ii) cavity with K/N=1/2, for instance confocal (iii)
cavity with K/N=2/3. K and N are integers. for (ii) and (iii) we
indicated besides each peak the first possible modes $(p,s)$. For
simplicity sake, we represented on (iii) the
 peaks corresponding to other combs on grey dashed line.
 }\label{peigne}
\end{center}
\end{figure}

Let us first consider the comb of cavity resonant lengths (see
figure (\ref{peigne})). Rewriting equation (\ref{numnp}) as:
\begin{equation}\label{numnp2}
L_{sp} =  \frac{\lambda}{2 N}\l( N p+ K (s+1)\r).
\end{equation}
where $p$ is an integer. One sees that whereas the free spectral
range of the cavity for longitudinal modes (p periodicity) remains
equal to the usual value $\lambda/2$, the natural unit to describe
the comb is $\frac{\lambda}{2N}$. N and K appear than as the
steps, in this natural unit, for the longitudinal comb (when
fixing s) and for the transverse comb (when fixing p). Within a
free spectral range, there exist N lengths for which the teeth of
the comb coincide, allowing us to define N \textit{families} of
modes.

Let us now consider the cavity in terms of rays optics.
(\ref{ABCDGouy}) implies that a paraxial cavity with a degeneracy
order $K/N$ verifies\cite{dingjan, Arnaut:69}:
\begin{equation}\label{ABCDn}
\left(T_{cav}\right)^N=\begin{pmatrix}
  A & B \\
  C & D
\end{pmatrix}^N
= \begin{pmatrix}
  1 & 0 \\
  0 & 1
\end{pmatrix} =  I_2 \mbox{.}
\end{equation}
where $I_2$ is the identity matrix of size $2 \times 2$. This
relation means that \textit{any} incoming ray will retrace back
onto itself after $N$ round-trips, forming a closed trajectory, or
\textit{orbit}. The total phase accumulated on such an orbit is $2
K \pi$  (as can be seen on equation (\ref{ABCDGouy}).

Up to now, only  perfect Fabry-Perot resonators have been
considered. If one consider a cavity with a given finesse
$\mathcal{F} \simeq \frac{2 \pi}{\gamma}$ where $\gamma$ is the
energy loss coefficient over one round-trip, supposed small, then
$\frac{\mathcal{F}}{2 \pi}$ is the mean number of round-trip of
the energy in the cavity before it escapes. As a consequence, for
a given finesse, and a cavity with a degeneracy order of $K/N$, we
have to compare $\mathcal{F}$ to $N$. If the finesse is low (i.e.
$\mathcal{F} \ll N$) then light will escape before retracing its
path and the previous discussion is not relevant. In the rest of
the discussion we will then stay in the high finesse limit
($\mathcal{F} \gg N$).

We now have all the tools necessary to study the propagation of a
paraxial image in a stable resonant cavity.

\section{Image transmission through a paraxial stable cavity}

\begin{figure}[htbp]
\begin{center}
\epsfig{file=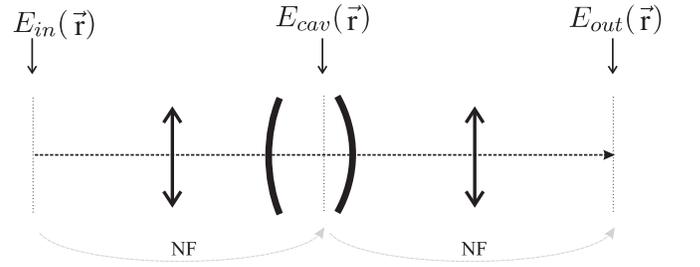,width=1\linewidth}
\end{center}
\caption{Scheme of the transmission of an image through a cavity.
 } \label{SchemaTransImageCavQCQ}
\end{figure}

We will consider for simplicity sake an impedance matched cavity,
where the input and output mirror have the same reflectivity and
no other losses exist in the cavity, so that at exact resonance a
mode is transmitted with an efficiency equal to unity. As shown on figure \ref{SchemaTransImageCavQCQ}, we define
the input image as the transverse field configuration
$E_{in}(\vec{r})$ at a chosen plane before the cavity. We want to
image it on a detection plane after the cavity. After propagation
along a first paraxial system corresponding to an ACT of
magnification equal to 1, $E_{in}(\vec{r})$ is transformed into
its near field at a given reference plane inside the cavity
(position $z_{ref}$). After propagation along a second identical
paraxial system, a new near field (output image) is obtained with
unity magnification after the cavity on a detection plane (a CCD
for instance). As the three planes are perfectly imaged on each
other, we will use the same notation $\vec{r}$ for these three
transverse coordinates, and we will omit the $z$ coordinate.

Let $a_{m,n}$ be the projection of the image on the mode $A_{m,n}$
of the cavity:
\begin{equation}
a_{m,n} = \int E_{in}\left(\vec{r}\right) A^*_{m,n}(\vec{r}) d^2
\vec{r}
\end{equation}

We can write $E_{in}$ as:
\begin{equation}
E_{in}\left(\vec{r}\right)= \sum_{m,n} a_{m,n} A_{m,n}(\vec{r}).
\end{equation}

The effect of the cavity on the image can be understood as a
complex transmission $t_{m,n}$ on each mode $A_{m,n}$, depending
on the length and geometry of the cavity. The output image will
then be written as:
\begin{equation}\label{ChampEoutpqGeneral}
E_{out}\left(\vec{r}\right)= \sum_{m,n} t_{m,n} a_{m,n}
A_{m,n}(\vec{r})
\end{equation}

\subsection{Single mode cavity}

Let us consider a single mode cavity having a length L chosen so
that only the $TEM_{00}$ resonates. The transmission function of
the cavity is :
\begin{equation}
t_{m,n}= \delta_{m 0 }\delta_{n 0 }
\end{equation}
and the output image is:
\begin{equation}
E_{out}\left(\vec{r},t\right)=  \sum_{m,n} t_{m,n} a_{m,n}
A_{m,n}(\vec{r}) = a_{0,0} A_{0,0}(\vec{r})
\end{equation}
All the transverse information about the input image $E_{in}$ is
then lost when passing through the cavity. In such a single
mode-cavity, the Gouy phase-shift $\alpha/2\pi$ is not a rational
fraction, so that whatever $N$, $T_{cav}^N \neq I_2$. In terms of
geometrical optics, this means that no ray (except the ray on the
optical axis) ever retraces its path on itself. This is the usual
understanding of the effect of a cavity on an image, where the
image is completely destroyed.

In general the precise length of the resonator is controlled
through a piezo-electric transducer. If the single-mode cavity
length is scanned over one free spectral range, every
Laguerre-Gauss cavity eigenmode will be transmitted one after the
other. The intensity field averaged over time on a CCD will be, at
a given transverse position :

\begin{equation}
<I_{out}(\vec{r})> \propto \sum_{m,n} |a_{m,n}| ^2
|u_{m,n}(\vec{r})|^2.
 \end{equation}

Each mode is transmitted at a different moment and does not
interfere with the others. As a consequence we obtain the sum of
the intensity into each $TEM_{pq}$ mode of the image, and not the
image since $\sum_{m,n} |a_{m,n}| ^2 |u_{m,n}(\vec{r})|^2 \neq
\left|\sum_{m,n} a_{m,n} u_{pq}(\vec{r})\right|^2$. This means
that, even scanned, a single-mode cavity does not transmit
correctly an image.

\subsection{Planar cavity}

It is important to study the planar cavity, since it is both
widely used experimentally and often taken as a model cavity in
theoretical works. Let us consider a planar cavity of length L.
The Gauss matrix is :
\begin{equation}\label{ABCDplan}
\begin{pmatrix}
  A & B \\
  C & D
\end{pmatrix}= \begin{pmatrix}
  1 & 2 L \\
  0 & 1
\end{pmatrix}
\end{equation}
It does not fulfill condition (\ref{ABCDn}) for any $N$ value, and
is therefore not degenerate. Strictly speaking, the planar cavity
is not a paraxial cavity, even for rays making a small angle
$\beta$ with the cavity axis, which escape from the axis after a
great number of reflections. As a consequence, there is no
gaussian basis adapted to this cavity. The planar cavity
eigenmodes are the tilted plane waves $e^{i k ( \beta_1 x +
\beta_2 y)}$, which are \textit{not degenerate} since they
resonate for different lengths: $L=p
\frac{\lambda}{2}(1+\beta_1^2/2+\beta_2^2/2)$. For a given length
the cavity selects a cone of plane waves with a given value of
$\beta_1^2+\beta_2^2$. The planar cavity is therefore \textit{not}
an imaging cavity. However, given a detail size, if the finesse is
low enough and the cavity short enough for the diffraction to be
negligible, then the image can be roughly transmitted. This study
is again outside the scope of this paper.

\subsection{Totally degenerate cavity}

Let us now consider a completely degenerate paraxial cavity, in
which all the transverse modes are resonant for the same cavity
length. As a consequence the transmission function of this cavity
brought to resonance is:
\begin{equation}
t_{m,n}= 1
\end{equation}
and the output field will be:
\begin{equation}
E_{out}\left(\vec{r}\right)=  \sum_{m,n} t_{m,n} a_{m,n}
u_{m,n}(\vec{r}) =E_{in}\left(\vec{r}\right).
 \end{equation}

Its Gauss matrix is  $T_{cav}=I_2$, its degeneracy order is 1:
\textit{every} input ray will retrace its path after a single
round-trip. A completely degenerate cavity can be called
self-imaging. Examples of self imaging cavities have been
described in\cite{Arnaut:69}.

\subsection{cavity of degeneracy order $K/N$}

Let us now study the propagation of an image through a transverse
degenerate cavity with degeneracy order $K/N$. We will use a
formalism of self-transform function, that we introduce in the
next subsection.

\subsubsection{Cyclic transforms}

Some functions are there own Fourier transform. They verify:

\begin{equation}
\tilde{f}(u)=f(u)
\end{equation}
where the Fourier transform $\tilde{f}$ is defined by:
\begin{equation}
\tilde{f}(u)=\int_{-\infty}^{+\infty}f(x)e^{2 \pi i ux } \dd x
\mbox{.}
\end{equation}

Two well known examples are the gaussian functions $f(x)= \alpha
e^{-\pi x^2}$ and the infinite dirac comb $f(x)=\sum_n \delta
(x-n)$. These functions are called  Self-Fourier functions or
SFFs. Caola\cite{Caola:91} showed that for any function $g(x)$,
then $f$ defined as:
\begin{equation}\label{eqCaola}
f(x)= g(x) + g(-x)+ \tilde{g}(x)+ \tilde{g}(-x)
\end{equation}
is a SFF.  Lohmann and Mendlovic\cite{Lohmann:92a} showed later
that this construction method for a SFF (equation (\ref{eqCaola}))
is not only sufficient but necessary. Any SFF $f(x)$ can be
generated through equation (\ref{eqCaola}) from another function
$g(x)$. Lipson \cite{Lipson:93} remarked that such distributions
should exist in the middle of a confocal resonator. Lohmann
\cite{Lohmann:94} also studied how such states could be used to
enhance the resolution in imaging.

It is straightforward to generalize this approach to a N-cyclic
transform. A transform $\mathcal{T}_C$ is said to be N-cyclic if
applied N times to \textit{any} function $F$ one gets the initial
function :
$$\mathcal{T}_C^N [F(x)]=F(x)$$
Let $\mathcal{T}$ be any transform. A function $F_S$ will be a
self-transform function of $\mathcal{T}$ if :
$$\mathcal{T} [F_S(x)]=F_S(x)$$
Given $\mathcal{T}_C$ a N-cyclic transform, and $g(x)$ a function,
it has been shown in \cite{Lohmann:92a} that $F_S(x)$ defined as:
\begin{equation}\label{generatingfunction}
F_S(x)=g(x)+\mathcal{T}_C[g(x)]+\mathcal{T}_C^2[g(x)]+...+
\mathcal{T}_C^{N-1}[g(x)]
\end{equation}
is a self-transform function of $\mathcal{T}_C$ and that any
self-transform function $F_S$ of $\mathcal{T}_C$ can be generated
in this manner (take $g=F_S/N$ for instance). The Fourier
transform is 4-cyclic. Other cyclic transforms and associated
self-transform functions are studied in
\cite{Patorski:89,Lipson:93,Lohmann:92b}.

We will show here that degenerate cavities produce such
self-transform functions from an input image through a
transformation similar to equation (\ref{generatingfunction}).

\subsubsection{Image propagation through a $K/N$ degenerate cavity}

Let us consider a resonator cavity with order of degeneracy $K/N$.
Let  $\gamma$ be the (low) intensity losses over one round trip on
the mirrors. For an impedance-matched cavity without internal
losses, the losses are identical on the two mirrors, meaning that
the amplitude transmission of one mirror is
$t=\sqrt{\frac{\gamma}{2}}$. For a cavity at resonance , we have:
\begin{equation}\label{TcavN}
\mathcal{T}_{cav}^N[E_{in}(\vec{r})]=E_{in}(\vec{r})
\end{equation}
since after N turns the field comes back onto itself.  It means we
can view $\mathcal{T}_{cav}$ as a $N$-cyclic transform on the
intensity.

The output field at resonance will be:
\begin{equation}\label{EoutcaviteKN}
E_{out}(\vec{r}) = t^2 \sum_{n=0}^{\infty}
\left[(1-t^2)\mathcal{T}_{cav}\right]^{n}E_{in}(\vec{r})
\end{equation}
the factor $(1-t^2)$ taking into account the double reflection of
the field at each round-trip. Using the fact that
$\mathcal{T}_{cav}^N[E_{in}(\vec{r})]= E_{in}(\vec{r})$, we can
finally rewrite the output field as :
\begin{eqnarray}\label{TransmCavKN}
E_{out}(\vec{r}) &=& \frac{t^2}{1-(1-t^2)^{N}}
\sum_{n=0}^{N-1}(1-t^2)^{n} \mathcal{T}_{cav}^n[E_{in}(\vec{r})]
\end{eqnarray}
In the high finesse limit ($\mathcal{F}\gg N$) $(1-t^2)^n\simeq 1-
n t^2 \simeq 1$ for $n\leq N$, so that:
\begin{eqnarray}
E_{out}(\vec{r}) \simeq \frac{1}{N} \sum_{n=0}^{N-1}
\mathcal{T}_{cav}^n[E_{in}(\vec{r})]
\end{eqnarray}
The output image is thus the self-transform field for $N$-cyclic
transform $\mathcal{T}_{cav}$, constructed from the input image
through the method of equation (\ref{generatingfunction}).

Let us finally note that most of this discussion can be extended
to more complex cavities, provided $\mathcal{T}_{cav}$ is a cyclic
transform, and that the present formalism holds for single-mode or
totally degenerate cavities: in the former case it means that a
self-transform function for a non-cyclic transform is just a
cavity mode; in the latter case the transform is just the
identity, and of course any field is a self-transform for
identity.

\section{Hemi-confocal cavity}

\begin{figure}[htbp]
\begin{center}
\epsfig{file=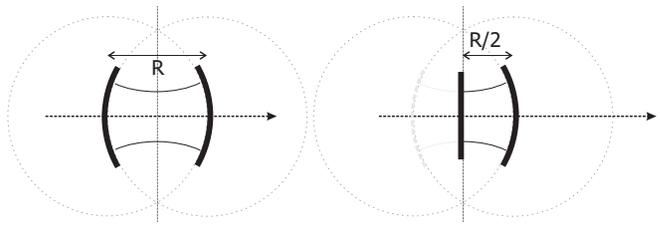,width=1\linewidth} \caption{The
confocal cavity (left) has a symetry plane. Placing a plane mirror
in this plane gives us the hemi-confocal cavity (right).}
\label{confsemiconf}
\end{center}
\end{figure}

We will now illustrate this formalism by considering in more
detail a particular cavity, the hemi-confocal cavity, which is
made of a plane mirror and a curved mirror R of radius of
curvature $R$ separated by a distance $L=R/2$ (see figure
\ref{confsemiconf}), which has already been studied in terms of
beam trajectories\cite{Chen:04}. We have studied this kind of
cavity both theoretically and experimentally in the framework of
our experimental investigations on cw amplification of images in a
cavity\cite{Gigan}.

\subsection {Theoretical study}

It is straightforward to show that the round-trip Gouy phase-shift
$\alpha$ is equal to $\pi/2$ for a hemi-confocal cavity, so that
its degeneracy order is $1/4$: there are four distinct families of
transverse modes, depending on the value $p + q$ modulo 4. The
round-trip Gauss matrix, starting from the plane mirror, is:
\begin{equation} \label{AR}
  T_{cav} =  \begin{pmatrix}
    0 & \frac{R}{2} \\
    - \frac{2}{R} & 0 \
  \end{pmatrix}
  \end{equation}
so that:
\begin{equation}
T_{cav}^2 = \begin{pmatrix}
    -1 & 0 \\
    0 & -1 \
  \end{pmatrix}
\mbox{,   } T_{cav}^4=\begin{pmatrix}
    1 & 0 \\
    0 & 1 \
  \end{pmatrix}
\end{equation}
So two round-trips give the opposite of the identity (symmetry
with respect to the cavity axis), which is the Gauss matrix of the
confocal cavity, and four round-trips give the identity, as
expected for a cavity with degeneracy order $1/4$.

$T_{cav}$ is the transformation of a f-f system, and is an exact
AFT transform:
\begin{equation}
\mathcal{T}_{cav} : u(\vec{r_1}) \rightarrow - e^{i k
L}\frac{2i}{R \lambda}\iint \dd^2 \vec{r}_1 u(\vec{r}_1)
\exp\left[{ -i \frac{ 4 \pi  }{\lambda R }}\vec{r}_1
\vec{r}_2\right]
\end{equation}
It is equal to the 2-D spatial Fourier transform, of the form:
\begin{equation}
{\tilde{u}}\left(\vec{y}\right)=\frac{2}{\lambda R }\int
u\left(\vec{r}\right)e^{-i \frac{4 \pi}{\lambda
R}\vec{y}\vec{r}}d^2\vec{r}
\end{equation}
multiplied by a phase factor  $a=i e^{i k L}$, which depends on
the exact length of the cavity. It must verify $a^{4} = 1$ at
resonance, so that $a = 1, i, -1$ or $ -i$.

\begin{figure}[htbp]
\begin{center}
\epsfig{file=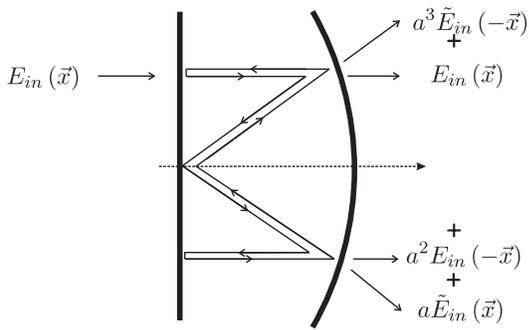,width=.8\linewidth} \caption{Ray
trajectory picture in the hemi-confocal cavity }
\label{resonnanceSC}
\end{center}
\end{figure}

If $E_{in}\left(\vec{r}\right)$ is the input image, then the
output field is (see figure (\ref{resonnanceSC})) :
\begin{equation}\label{ChampSCSortieGeneral}
  E\left(\vec{r}\right) = \frac{1}{4} \l(E_{in}\left(\vec{r}\right)+ a^2 E_{in}\left(-\vec{r}\right)\r)
  + a \l(\tilde{E}_{in}\left(\vec{r}\right)+ a^2 \tilde{E}_{in}\left(-\vec{r}\right)\r)
\end{equation}

In terms of imaging, $a$ is the phase between then even/odd parts
of the field and its spatial Fourier transform, and $a^2$ gives
the parity of the output image. Each value of $a$ corresponds to a
given family of modes, more precisely:
\begin{equation} \label{FamillesSC}
  \begin{cases}
    a = 1 & \longrightarrow\text{modes } m+n= 0[4] \\
   a = i & \longrightarrow\text{modes } m+n= 1[4] \\
   a = -1 & \longrightarrow\text{modes } m+n= 2[4] \\
   a = -i & \longrightarrow\text{modes } m+n= 3[4]
  \end{cases}
\end{equation}

For example, the hemi-confocal cavity tuned on the $m+n= 0[4]$
family will transmit the sum of the even part of the image and of
the even part of its Fourier transform.

\subsection{Numerical simulation}

\begin{figure}[htbp]
\begin{center}
\epsfig{file=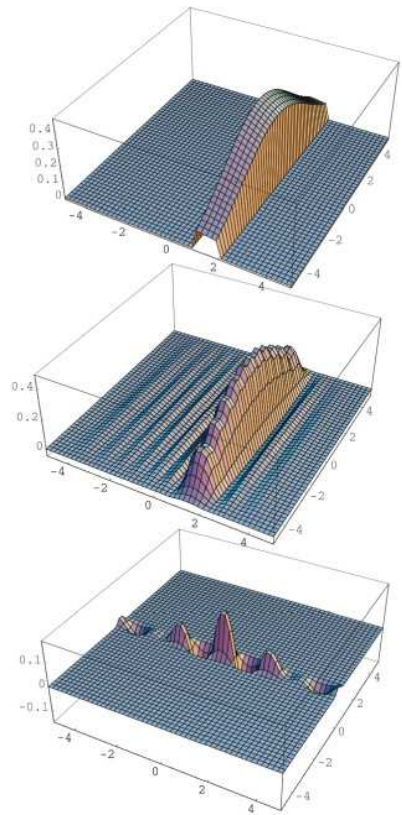,width=0.6\linewidth}
\end{center}
\caption{Input image: infinite slit intercepting a large gaussian
mode (up), projection on the first 400 modes of the cavity
(middle), and spatial Fourier transform (down). }
\label{FenteGaussienneIn}
\end{figure}

We will now give results of a numerical simulation in a simple
experimental configuration: in order to create the input image
$E_{in}$, a large gaussian $TEM_{00}$ mode is intercepted by a
single infinite slit of width $w_0$, shifted from the optical axis
by $1.5 w_0$, which is imaged (near field) onto the reference
plane ($z_{ref}$) of the cavity. Without the slit, the input
$TEM_{00}$ mode has in the reference plane a size equal to three
times the waist of the $TEM_{00}$ cavity eigenmode. We study the
transmission of this input image through the cavity at the near
field detection plane. We represented on figure
(\ref{FenteGaussienneIn}) the input image, its decomposition over
the first  400 $TEM_{pq}$ modes, (with $ 0 < p, q < 20$), and its
spatial Fourier transform.  Limiting the decomposition of the
image to only $\sim 400$ modes is equivalent to cutting high-order
spatial frequencies, and therefore takes into account the limited
transverse size of the optics.

\begin{figure}[htbp]
\begin{center}
\epsfig{file=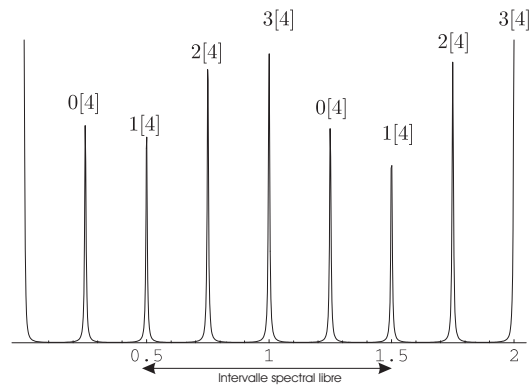,width=.8\linewidth}
\end{center}
\caption{simulation of the transmissions peaks of the cavity for
the slit of figure (\ref{SchemaTransImageSC}), for a finesse
$\mathcal{F}=500$. } \label{PicsFenteTheorique}
\end{figure}

Figure \ref{PicsFenteTheorique} gives the expected transmission
peak as a function of cavity length, and displays the four
families of modes in a given free spectral range. The height of
each peak is proportional to the intensity of the image projection
on the family of mode. For instance a symmetric field will have no
intensity on the $1[4]$ and $3[4]$ families.

\begin{figure}[htbp]
\begin{center}
\epsfig{file=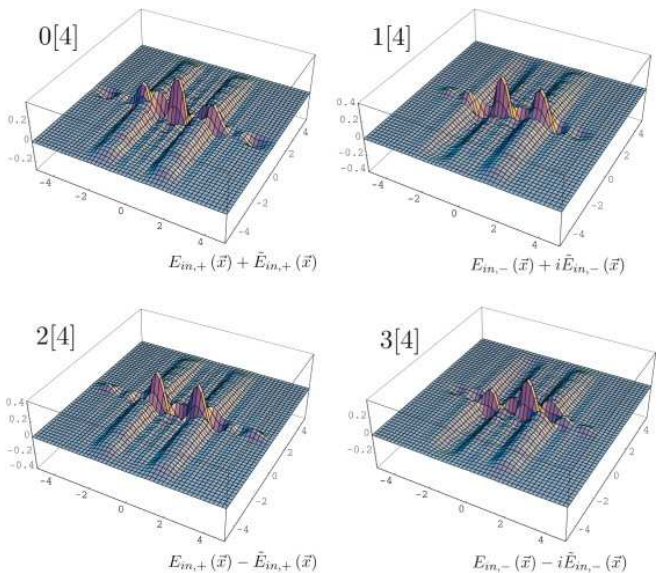,width=1\linewidth}
\end{center}
\caption{Theoretical transmission (amplitude) of the slit by the
hemi-confocal cavity, for every mode family $m+n=i[4]$. }
\label{TransmissionSCpic1234}
\end{figure}

Figure \ref{TransmissionSCpic1234} gives the amplitude of the
transmitted field for each family of modes, calculated from the
transmission of the 400 first $TEM_{mn}$ modes. For each family,
one easily recognizes the even or odd part of the image (two
slits) and the Fourier transform along the axis perpendicular to
the slits.

\begin{figure}[htbp]
\begin{center}
\epsfig{file=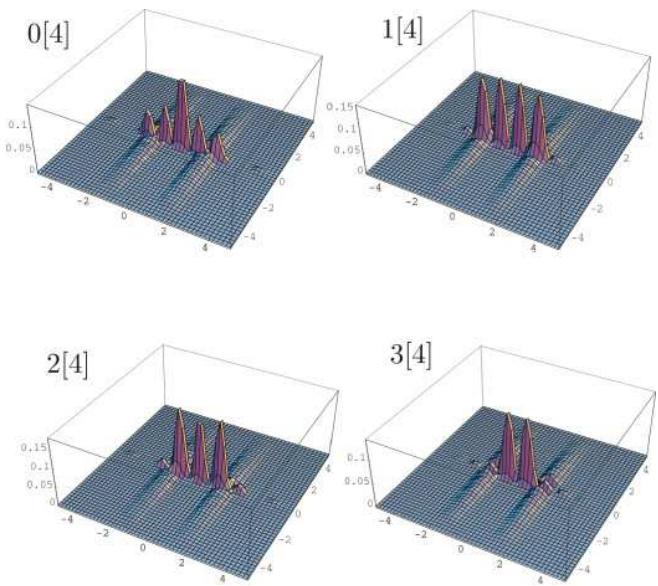,width=1\linewidth}
\end{center}
\caption{Theoretical transmission (in intensity) of the slit by
the hemi-confocal cavity, for every mode family $m+n=i[4]$. }
\label{TransmissionSCInt1234}
\end{figure}

The expected intensity image is represented on figure
\ref{TransmissionSCInt1234} for each family. One observes that
the Fourier transform is much more intense that the transmitted
image, even though equation (\ref{ChampSCSortieGeneral}) shows
that there is as much energy in the Fourier transform than in the
image. In the present case, the Fourier transform is much more
concentrated than the image, which is the reason why the local
intensity is higher. As the parity information on the field
disappears when looking at the intensity, it is difficult to infer
from it which resonance is involved. An indication can come from
the intensity on the optical axis, which is always zero for an
antisymmetric output. One can note that if we add the amplitude
fields corresponding to the resonances $m+n=j[4]$ and
$m+n=j+2[4]$, the two terms corresponding to the Fourier transform
vanish. One only gets the even or odd part of the image, which
corresponds to the action on the image of a confocal cavity. It is
interesting to note that no combination of modes transmit only the
Fourier transform of the image.

An interesting question is to know which information is lost in
the transmission through the cavity, since one only transmits a
quarter of the input intensity. By looking at the transmitted
image, it seems that no information is really lost, except on
areas where the image and its Fourier transform overlap. But here
we have some a priori information on the image that we have sent
and we know which part of the output is the image, and which part
is the Fourier transform. In more general cases, the information
we lose is the knowledge about whether what we observe is the
image or the Fourier transform, as well as half the image (since
the parity is fixed by the geometry of the cavity, only half the
output image is relevant to reconstruct it). Therefore for a given
resonance, this cavity not only cuts 75$\%$ of the modes, it also
destroys 75 $\%$ of the information.

As a conclusion, the transmission through the cavity transforms
the input image into its its self-transform function corresponding
to the round-trip transform of the hemi-confocal cavity. One may
notice that for the resonance $m+n=0[4]$, a self-Fourier image,
i.e a SFF field, is obtained.

\subsection{Experimental demonstration}

\begin{figure}[htbp]
\begin{center}
\epsfig{file=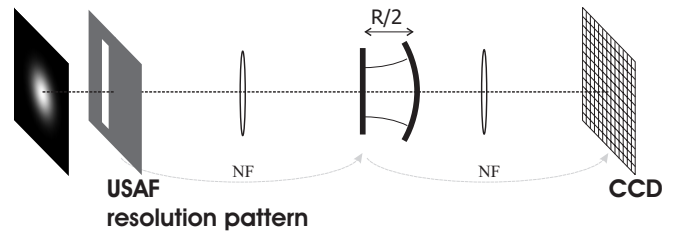,width=1\linewidth}
\end{center}
\caption{Schematic representation of the experimental setup.}
\label{SchemaTransImageSC}
\end{figure}

\begin{figure}[htbp]
\begin{center}
\epsfig{file=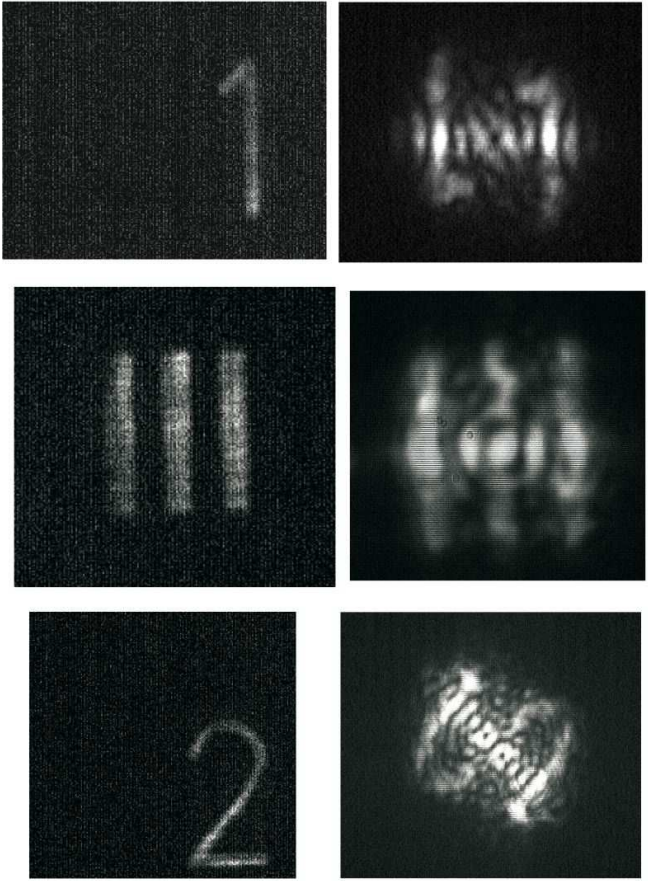,width=1\linewidth}
\end{center}
\caption{Image on the resolution target (left) and their
transmission through the hemi-confocal cavity (right). }
\label{TransImExpSC}
\end{figure}

For practical reasons, we had to use a hemi-confocal cavity in our
experimental set-up designed to study parametric image
amplification in optical cavities (see figure
\ref{SchemaTransImageSC}). We placed a USAF (US Air Force)
resolution target on the path of a $TEM_{00}$ mode produced by a
Nd:YAG laser, and imaged it onto the plane mirror of a
hemi-confocal cavity, of length $50mm$, servo-locked on a
resonance peak. The size of the $TEM_{00}$ mode inside the cavity
was three times larger than the eigenmode waist of the cavity. The
finesse of the cavity was about 600. The plane mirror of the
cavity was then imaged on a CCD camera. The experimental
transmitted images, together with the corresponding objects, are
represented on figure (\ref{TransImExpSC}). The size of the
$TEM_{00}$ cavity mode is roughly equal to the width of the
transmitted slit in the second line. One notices that each output
image is symmetric, the center of symmetry being the axis of the
cavity. It is possible to recognize on the transmitted images the
symmetrized input image, and the patterns at the center corresponding
to the Fourier transform of the input. For a slit it is well known
that its Fourier transform is the sinc-squared diffraction
pattern, perpendicular to the slit. This kind of pattern can be
recognized on the upper two images of the figure. On the last
image the symmetrized "2" is somewhat truncated by the limited
field of view imposed by the size of the illuminating
$TEM_{00}$ mode, whereas the diffraction pattern has the general
shape of the Fourier transform of the slit formed by the main bar
of the "2", tilted at 45$^\circ$, plus a more complex shape
corresponding to the Fourier transform of the remaining part of
the image.

\section{Conclusion}

In summary, this paper has studied in a general way the problem of
image transmission through a paraxial cavity, characterized by its
round-trip (ABCD) matrix, the eigenvalues of which give the
round-trip Gouy phase shift, and therefore the order of transverse
degeneracy of the cavity. We have shown that the formalism of
self-transform functions, already applied in optics but never to
cavities, was very useful to understand how an image is
transmitted through a degenerate cavity: at resonance the cavity
transmits the self-transform part of the input field. We have then
focused our attention on the hemi-confocal cavity, which performs
a spatial Fourier transform over one round trip, and shown that it
transmit the self-Fourier part of the image. This property was
demonstrated experimentally on various shapes of input images.
Furthermore we have shown that a transverse degenerate cavity is a
very convenient way to produce a self-transform field from any
input field, for instance in the case of the hemi-confocal cavity
a field which is its own Fourier transform, i.e. its own
far-field). Such states are interesting for
optics\cite{Lohmann:94} and in physics in general\cite{Lipson:93}.

From a more practical point of view, transverse degenerate
cavities can be useful for imaging purposes. For example they are
necessary for intracavity c.w. parametric amplification of images.
The observation of c.w. image amplification with low pump powers
will be reported in a forthcoming publication \cite{Gigan}. These
experimental results open the way to the observation of specific
quantum aspects of imaging which have been predicted to occur in
such devices, such as noiseless phase-sensitive amplification,
local squeezing or spatial entanglement.

\appendix*

\section{Eigenvectors and Gouy phase of a cavity}

Let A,B,C and D be the coefficients of the cavity round-trip Gauss
matrix $T_{cav}$, starting from any plane. Given that $AD-BC=1$,
the eigenvalues of this Gauss are:
\begin{equation}\label{q4}
\mu_{1,2} = e^{\pm  i \arccos \frac{A+D}{2}}
\end{equation}
They are simply related to the matrix trace $A+D$, and as expected
independent of the reference plane one choses in the cavity to
calculate the Gauss matrix. Let us now consider the fundamental
gaussian mode of the cavity, $E(r)=E(0) e^{-ik r^2/2q}$, where $q$
is the complex radius of curvature. Using the propagation relation
(\ref{HuygensParaxial}), one easily computes that, on axis, it
becomes after one round trip:
\begin{equation}
E'(0)=E(0) e^{i k L} \frac{1}{B/q + A}
\end{equation}
The round-trip Gouy phase shift $\alpha$ for this mode is
therefore:
\begin{equation}\label{q2}
\alpha=Arg [\frac{1}{B/q + A}].
\end{equation}

On the other hand, after one round trip in the cavity, the complex
radius of curvature becomes $\frac{A q +B}{C q +d}$, but since it
is an eigenmode of the cavity,  $q$ must verify :
\begin{equation}\label{q1}
q=\frac{A q +B}{C q +d}
\end{equation}
from which one deduces:
\begin{equation}\label{q3}
A+\frac{B}{q}=\frac{D+A}{2}+i\sqrt{1-(A+D)^2/4}
\end{equation}
From equation(\ref{q2}), one then find that $\alpha = arcos \left(
\frac{A+D}{2}\right)$, and therefore using Eq(\ref{q4}) one
retrieves relation(\ref{ABCDGouy}).

\begin{acknowledgments}

Laboratoire Kastler-Brossel, of the Ecole Normale Sup\'{e}rieure
and the Universit\'{e} Pierre et Marie Curie, is associated with
the Centre National de la Recherche Scientifique.\\
This work was supported by the European Commission in the frame of
the QUANTIM project (IST-2000-26019).

\end{acknowledgments}

\end{document}